\documentclass{llncs}
\usepackage{graphicx}
\usepackage{amsmath}
\usepackage{subfig}

\begin{document}

\title{Net Stable Funding Ratio: Impact on Funding Value Adjustment }
\titlerunning{Hamiltonian Mechanics}  
%
\author{Medya Siadat\inst{1} \and Ola Hammarlid\inst{2} 
}
%
%
%
\institute{SEB, Stockholm, Sweden\\
\email{medya.siadat@seb.se}
\and
Swedbank, Stockholm, Sweden\\
\email{Ola.Hammarlid@swedbank.se}
}
\maketitle              

\begin{abstract}
In this paper we investigate the relationship between Funding Value Adjustment (FVA) and Net Stable Funding Ratio (NSFR).  FVA is defined in a consistent way with NSFR such that the new framework of FVA monitors the costs due to keeping NSFR at an acceptable level, as well.
 In addition, the problem of choosing the optimal funding strategy is formulated  as a shortest path problem where the proposed FVA framework is applied in the optimization process. The solution provides us with the optimal funding decisions that lead to the minimum funding cost of the transaction. We also provide numerical experiments for FVA calculation and optimization problem. 
\end{abstract}
\section{Introduction}

The Basel Committee on Banking Supervision (BCBS) introduced the \emph{Net Stable Funding Ratio (NSFR)} in 2010 and the final standards were published in October 2014. NSFR is a measure designed to compare a firm's available stable funding (ASF) to its required stable funding (RSF). 
NSFR is defined  based on the liquidity characteristics of the firm's assets and activities and the aim is to ensure that the firm holds a minimum amount of stable funding over a one year horizon. 
NSFR will become a minimum standard by $1^{st}$ January 2018. So
Banks and other financial institutions will be required to 
take it into account to reduce funding risks originating from lack of liquidity. In fact NSFR is a complementary   for Liquidity Coverage Ratio (LCR) which is a short term  measure that  requires banks to hold enough high quality  liquid assets  in order to survive under stressed conditions within 30 days time horizon.   

The NSFR is defined as the amount of available stable funding relative to the amount of required stable funding:
\begin{equation}\label{nsfrformula}
NSFR=\frac{ASF}{RSF} 
\end{equation}
 which should always be equal or greater than 1. 
 
\begin{figure}
  \centering
  {\includegraphics[width=0.45\textwidth]{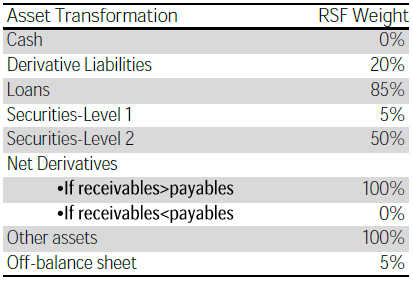}\label{rsf}}
  \hfill
  {\includegraphics[width=0.45\textwidth]{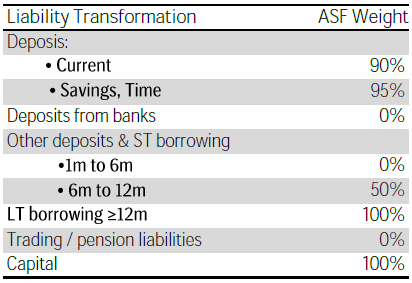}\label{asf}}
  \caption{Summary of transformation of assets and liabilities to NSFR }
  \label{weights}
\end{figure}

 ASF refers to those types of equity and liability that are supposed to provide stable sources of funding over one year time horizon. ASF is calculated by first assigning the carrying value of a firm's equity and liability to one of the specified categories and then multiply each assigned value by a weight which is already defined in NSFR standards (see Figure \ref{weights}). The total ASF is the sum of weighted amounts.

 RSF, on the other hand, measures different assets in terms of the proportion of stable funding required to support them. Similar to the ASF case, the total RSF is calculated as the sum of the value of assets multiplied by a specific weigh assigned to each particular asset type. Figure \ref{weights} shows ASF and RSF weights for different categories of assets and liabilities.

The idea behind introducing  NSFR is to restrict
over-reliance on short-term  funding and promote banks to have better assessment of liquidity risk across all on- and off-balance sheet items (see \cite{basel} for more details).

\subsection*{ Funding Value Adjustment (FVA)}
 FVA  mostly reflects the costs of collateral that banks post to hedge  uncollaterallized derivatives. This cost increased rapidly after financial crisis and this raise encouraged  banks to pay  attention to funding valuation adjustment in addition to CVA and DVA. 

Different structures for FVA formulation has been discussed so far (see \cite{hull1,bur1,bur2,albanese1,albanese2,albanese3,albanese4}),for example, 
equation \eqref{fva_albanese} refers to FCA (Funding Cost Adjustment) introduced in \cite{albanese1}:
 
 \begin{equation}\label{fva_albanese}
 FCA=E\Big[\int_0^{\infty}e^{-\int_{0}^{t}r_{\text{B}}(s)ds}s_B(t)\sum_{i}V_i^{+}(t)dt\Big]
 \end{equation}
 where $E$ denotes risk-neutral expectation, $V_i(t)$ is the e value of the $i^{th}$ unsecured
netting set, $r_B(t)$ is the bank's funding rate defined as $r_B(t) = s_B(t) + r(t)$  and $s_B$ is the
bank's funding spread over the LIBOR rate.
We see  that the positive value of portfolio incorporates the funding cost and the possibility
of bank default is ignored.  The implementation of the above methodology is done in section \ref{numex} and the results are compared to our proposed approach.
 

In a very simplified manner, a derivative desk of a financial institution, e.g. a bank, sells derivative securities to clients while hedging them with other dealers. This operations bring a cost to the bank which can be provided by different sources (in most cases by a debt from the market). 
These operations  impact the bank's liquidity profile, measured by NSFR.  
We would like to define FVA such that it monitors the NSFR costs, as well.

Also the problem of choosing the best funding strategy at each time step (within the lifetime of the contract) is disscussed. In fact, we show that this problem leads to an optimization problem which can be formulated by a shortest path problem. By solving this problem one can obtain the funding strategy that leads to minimum funding cost.

\begin{figure}
\centering
\includegraphics[width=\textwidth]{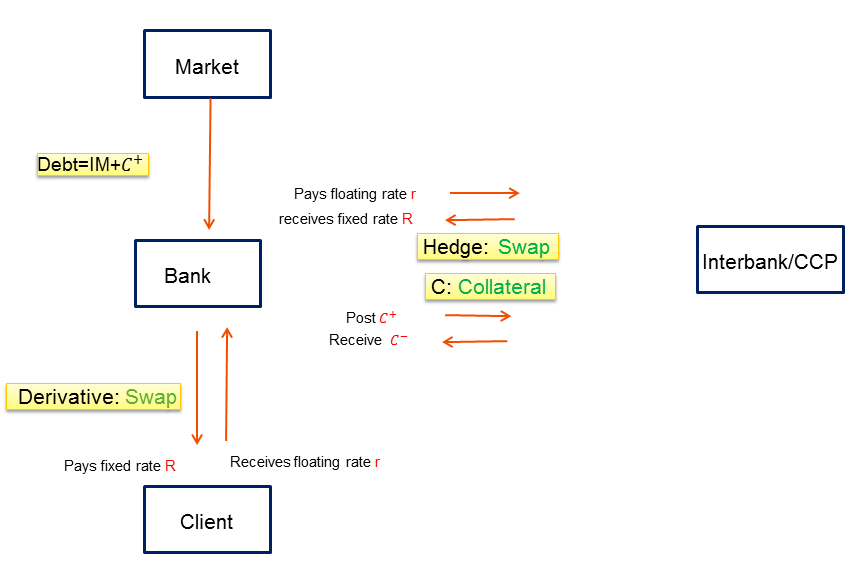}
\caption{Toy model containing a single derivative with a client (no CSA)  which is hedged with  the interbank in the presence of collateral.}\label{model}
\end{figure}

We consider a simple model  to establish  our framework (see Figure \ref{model}). Suppose that bank enters into an interest rate swap with a client without CSA and hedge the OTC by a back-to-back swap with interbank which is collateralized. The required fund for posted collateral and initial margin is borrowed from market.Next section describes the mechanism of FVA for this toy model, also we discuss the structure of optimization problem in more details. 

\section{Methodology}
Consider  the model described in Figure \ref{model}, the NSFR calculation for this model can be easily done using the table in Figure \ref{weights}. The ASF calculation includes multiplying regulatory capital (Reg.Cap) and the debt (D) by their associated weights, so 
\[ASF=100\% \text{Reg.Cap} + \alpha D\]
where 
\[ \alpha=
\begin{cases}
50 \% & \text{Debt with 6 months--12 months maturity}\\
100 \% & \text{ Debt with more than 1 year maturity}
\end{cases}
\]
Also the RSF can be calculated in the same way:
\begin{equation*}\label{rsf}
RSF =100 \%(\text{Net Derivatives - Net Collateral})+20\%(\text{Derivative Liabilities})
\end{equation*}
and then the value of NSFR is obtained from $\text{NSFR}=\dfrac{\text{ASF}}{\text{RSF}}$. 

Now the objective is to manage NSFR such that it remains at the desired level. It can be seen from the ASF and RSF calculations that the only viable modification that one can do is to adjust $D$,  both in terms of amount and the maturity time and the aim is to keep the NSFR level at 1. At this stage we assume that  \( \alpha\) is fixed during the life time of the swap which means that the maturity of the debt is considered to be same for all payment dates. Then the amount of debt can be easily calculated by forcing NSFR to be equal to $1$:
\begin{equation}
1=\frac{ASF}{RSF}=\frac{100\% \text{Reg.Cap} + \alpha D}{\text{RSF}}
\end{equation}
So we can find $D$ in the above equation as :
\begin{equation}\label{NewD}
D=\frac{\text{RSF}-100\%  \text{Reg.Cap}}{\alpha}
\end{equation}

The core idea of the  proposed methodology is that costs of NSFR should be included in the total  funding cost of a transaction. So we define the new formula as : 
\begin{equation}\label{fvaNSFR}
 FVA=FVA_1+FVA_2
 \end{equation}
 where $FVA_1$ refers to the costs due to keeping NSFR at $1$, and $FVA_2$ shows the funding costs due to the posted collateral: 
 \begin{eqnarray}
FVA_1&=&E\Big[\int_0^{T}e^{-\int_{0}^{t}r_{B}(s)ds}s_B(t)D(t)dt\Big]\label{fva1} \\
FVA_2&=&E\Big[\int_0^{T}e^{-\int_{0}^{t}r_{B}(s)ds}\tilde{s}_B(t)\Big(C^+(t)-D(t)\Big)^+dt\Big]\label{fva2}
\end{eqnarray} 
It can be seen that when
$D\geq C^+$  (the debt borrowed from the market covers the posted collateral) then $FVA_2$ is disappeared from the total FVA. 

In order to calculate (\ref{fva1}) and (\ref{fva2})  we need to compute $D$ at each time step. Also the maturity of the debt at each payment date should be specified. The amount of $D$ is obtained from (\ref{NewD}) while the maturity time of the debt is a decision made by bank. So the problem of choosing the best  debt's maturity  is so important in FVA calculations and as we will show in the following, it leads to an optimization problem. 
 
Now we briefly introduce the dynamic programming structure. Here we only consider discrete-time models and explain how it is  applied in our case, for more technical details see \cite{DP1} and \cite{MCbook}.

Dynamic Programming (DP) is a powerful optimization principle which is widely used in a variety of problems. One of the applications of DP is to find the shortest path in a network. A network consists of a set of nodes and a set of arcs joining pairs of nodes. Each arc is labeled by a number that can be interpreted as the arc length (cost). The purpose is to find the path between the first node and the terminal node that has the total minimum length. Running a Monte-Carlo simulation for each possible path provides a good estimation of the expected length of the path. 
\subsubsection{The Shortest Path Problem}
Suppose that $V_i$ shows the length of the shortest path from node $i$ to the terminal node, if $\mathcal{I}$ shows the set of all node indexes then
\begin{equation}\label{nod1}
V_i=\min_{i,j}\{C_{ij}+V_j\},\quad \quad \forall j \in \mathcal{I}
\end{equation}
where $C_{ij}$ is the cost of going from node $i$ to all its immediate successors $j$ and $V_j$ is the optimal cost of going from  $j$ to the terminal node.
The above equation is a recursive problem with the boundary condition of $V_N=0$, where $N$ is the index of the last node. 

The goal is to find the value function $V_0$ for the initial node which is obtained by further decomposing of $V_j$. 

Lets $s_t$ shows the state variable  and $x_t$ shows the decision variable, then the cost is a function of the decision $x_t$ made at state $s_t$ (shown by  $C(s_t,x_t)$ in \eqref{nod2}).
The problem of finding the shortest path (minimum cost) is formulated as follows:
\begin{equation}\label{nod2}
V_0=\min E[\sum_{t=0}^{T} C(s_t,x_t)]
\end{equation}

In other words, making the decision $x_t$ at state $s_t$ causes the cost $ C(s_t,x_t)$ and we would like to minimize the expected cost incurred over a given planning horizon which is done by going backward from the terminal node.

\section{Numerical Experiments}\label{numex}
Suppose that the derivative in Figure \ref{model} is an interest rate swap where the bank agrees to pay  $2 \%$ per annum (with semiannual compounding) and receives 6-month LIBOR on a notional principle of $\$100$ million, with the maturty of $T=5$ years. 

In this example  the LIBOR rate $r_t$ is simulated by Vasicek interest rate model starting from \emph{OIS} rate, also  $r_B$ shows the bank's funding rate where $r_B=r+s_B$.
The calculation for valuing the swap in terms of bonds is done using the method explained in \cite{hullbook}, chapter 7.



 In order to calculate  NSFR measure, we need to first calculate ASF and RSF for all simulated paths and all time steps. Then  divide them for each simulated path and all time steps to get the NSFR values of that specific path. Finally the expected NSFR values are obtained by  averaging over all the simulated paths at each time step. In order to prevent the NSFR values to become infinity (or a very large value) we need to define a boundary for RSF. In this example  we suppose $\text{Derivative Liabilities}= 10000$ whenever it 
 goes below $\$ 10000$. 
 
The debt applied in ASF calculation is obtained from  the following  standard assumption:
\begin{equation}\label{debt}
\text{Debt}=\text{initial margin}+\text{posted collateral}
\end{equation}

\begin{figure}
  \centering
  \subfloat[Expected value of NSFR.]{\includegraphics[width=0.4\textwidth]{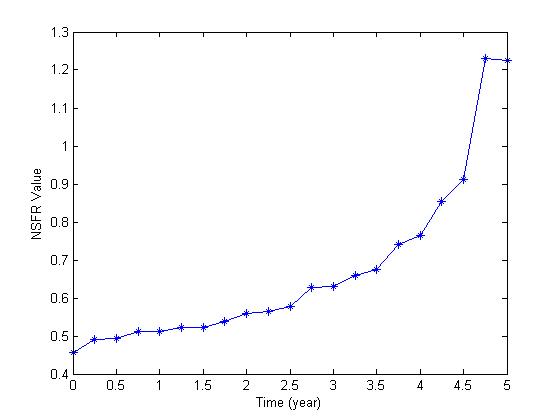}\label{a}}
  \hfill
  \subfloat[Heatmap of NSFR distribution.]{\includegraphics[width=0.55\textwidth]{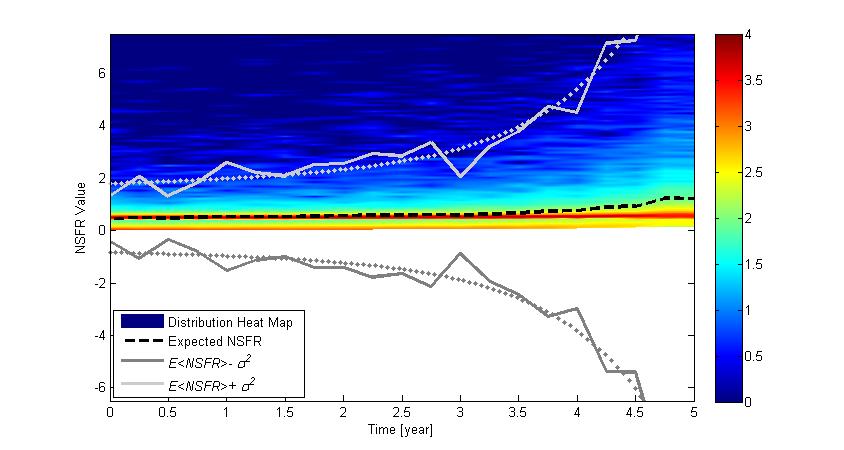}\label{b}}
  \caption{NSFR measure }
  \label{nsfrPLOT}
\end{figure}

 

Figure \ref{a} shows variations of the expected NSFR value during the life time of the derivative where the debt in ASF calculation is obtained from the standard approach of \eqref{debt}. It can be seen that NSFR is increasing gradually from around $0.5 $ and finally reaches above one in the last six month of the swap. 
Whenever NSFR is less than one, bank needs more debt in order to raise its level. The total amount of debt required to both increase the level of NSFR and to cover the posted collateral has been defined in \eqref{NewD}. We use this new amount of debt to calculate FVA from our proposed method in \eqref{fvaNSFR}.

Also Figure \ref{b} provides statistical information regarding the distribution of NSFR. The heatmap reveals the bimodal nature of the NSFR distribution with the dominant peak close to the expected value and a smaller peak close to zero. When the time increases, the distribution of NSFR becomes heavy-tailed as such the expected NSFR (dashed-black line) takes some distance from the dominant peak. This is due to enhancement of variance in each time step.The exponential growth of the variance of NSFR fluctuation with time is obvious from the Figure \ref{b} (\emph{cf.}\ solid gray lines and their exponential fit as dotted gray  lines)

\begin{figure}
\centering
\includegraphics[width=.7\linewidth]{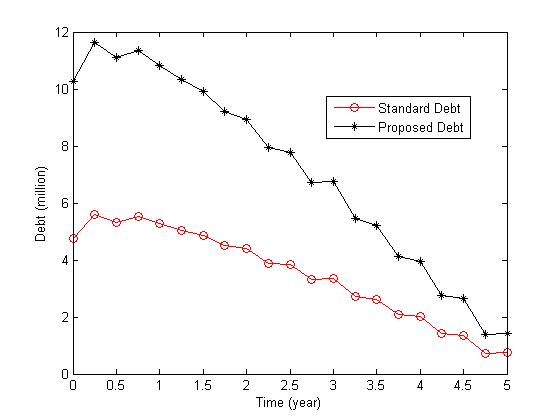}
\caption{ A comparison between the expected debt  at each time step calculated by the standard method in \eqref{debt} and the proposed approach of \eqref{NewD}. We assumed $r_{OIS}=1\%$ and $s_B=0.51\%$.}\label{twoDebt}
\end{figure}

 In the following we provide a comparison between  FVA values obtained from equation \eqref{fva_albanese} and our proposed approach in  \eqref{fvaNSFR} (see Table \ref{comparison}). We assume spread is $0.51\%$ so the bank's funding rate is $r_B=r+0.0051$, also all model parameters are assumed to be equal for both  methods. It can be seen  that  our proposed approach in \eqref{fvaNSFR} produces  higher values of FVA. The reason is that keeping the level of NSFR at $1$ or above causes some additional costs (NSFR cost) so the bank needs to increase the amount of debt. Figure \ref{twoDebt} shows the difference between the  amounts of expected debt when it is calculated from the standard approach of \eqref{debt} and when it is defined as we proposed in \eqref{NewD}. The graphs emphasize that the bank needs more debt when the NSFR costs are included in the total debt calculations. 
As a result including NSFR cost in the 
 FVA calculation framework leads to a higher overall funding cost.
\begin{table}
\centering
\begin{tabular}{|l||r|r|r|}
	\hline
   
	$\quad \quad  r_{OIS} \quad$ & $0.5\% \quad$  & $1\% \quad$ & $1.5\% \quad$ \\
    \hline
    
    Proposed  FVA in \eqref{fvaNSFR} & $0.1636$ &$0.1668$ & $0.1698$\\
	\hline
    
   FCA & $ 0.0799$ & $0.0802$& $0.0808$\\
	\hline
    
\end{tabular}

\caption{A comparison between the funding costs (million) obtained from \eqref{fva_albanese} and our proposed method in \eqref{fvaNSFR}. The spread is fixed at $0.51\%$ and the OIS rate is changing.}\label{comparison}
\end{table}

Taking a closer look at Figure \ref{twoDebt} and comparing the numbers of standard debt  and the modified debt reveals that in this example the modified debt is almost two times bigger than the standard debt.
   



In another experiment (Table \ref{tab2}), we supposed $r_{OIS}=0.5\%$  and  checked the variations of our proposed FVA  with different spreads. Here we assume  $s_B\leq0.5\%$ refers to the debts with the maturity between 6 months to 1 year while $s_B>0.5\%$ corresponds to the debts with the maturity of more than $1$ year. It can be seen in Table \ref{tab2}  that although FVA is gradually increasing with the growth of spread, using debts with the maturity of more than 1 year, decreases the funding costs of the contract.
\begin{table}
\centering
\begin{tabular}{|l||r|r|r||r|r|r|}
	\hline
   
	$\quad \quad  s_B \quad$ & $0.3\% \quad$  & $0.4\% \quad$ & $0.5\% \quad$  & $0.6\% \quad$  & $0.7\% \quad$ & $0.8\% \quad$ \\
    \hline
    
    Proposed  FVA in \eqref{fvaNSFR} & $0.1051$ &$0.1336$ & $0.1523$  & $0.1047$  & $0.1140$ & $0.1342 \quad$\\
	\hline
    
\end{tabular}

\caption{ FVA values (million) obtained from  our proposed method in \eqref{fvaNSFR} when $r_{OIS}=1\%$ is fixed  and the spread is changing.}\label{tab2}
\end{table}




\subsection{Optimization process}
In this section we try to formulate our optimization problem as a \emph{shortest path problem} described earlier, then we implement the dynamic programming algorithm to solve it.

The amount of debt required by bank is calculated from equation \eqref{NewD}  which  obviously depends on the parameter $\alpha$. 
On the other hand $\alpha$ is determined by the maturity of the debt so the decision of choosing the best maturity for the debt is of great importance as it directly affects the funding costs of the transaction. Since choosing different maturity dates leads to different funding costs,  the shortest path problem can be addressed as finding  the optimal funding strategy (maturity time ) at each payment date that provides us with the  minimum total funding cost.

In our case the state variable is represented by the simulated interest rate and the decision variable is set as the maturity of the debt. 
In this example  we assume that the costs can be funded via a debt with the  maturity of 6 months, 1 year or 2 years. So there are  three decision variables at each payment date.
Suppose that $r_{OIS}=1\%$ and the spreads for a 6 month, 1 year and 2 years debt are $s_{B_1}=0.5\%, s_{B_2}=0.51\%$ and $s_{B_3}=0.52\%$ , respectively. 
Then bank needs  to decide among one of the following rates 
at each payment date:
\begin{eqnarray*}
r_{B_1}&=&r+s_{B_1},\\
r_{B_2}&=&r+ s_{B_2}, \\ 
r_{B_3}&=&r+s_{B_3}.
\end{eqnarray*}

Also the cost function $C(s_t,x_t)$ of minimization problem  \eqref{nod2} can be interpreted as the funding cost of the transaction defined in equation \eqref{fvaNSFR}.


Applying dynamic programming method on the above problem reveals that the minimum funding value adjustment is obtained when the debt with 2 years maturity is chosen for all payment dates. 
Choosing the funding strategies in this way gives us 
$$ \text{Optimal FVA}= 0.1044 \quad \text{million}$$
 while the funding costs calculated from \eqref{fvaNSFR} with 1 year maturity debt for all time steps is $$\text{FVA}=0.1668\quad \text{million}.$$
 Comparing these values shows that  the optimal FVA is smaller than the normal FVA  but still greater than FCA obtained from equation \eqref{fva_albanese} at which no NSFR cost was included in the calculations:
 $$\text{ FCA}=0.0802\quad \text{million}$$

The strategy explained above can be easily extended to the whole portfolio where the netting value of all contracts is included in the calculations instead of a single transaction's value. Also the decision variables in the optimization process  can be increased to finitely many cases including more complex ones.



%
%
%

\end{document}